\documentclass[sigconf]{acmart}

\AtBeginDocument{%
  \providecommand\BibTeX{{%
    \normalfont B\kern-0.5em{\scshape i\kern-0.25em b}\kern-0.8em\TeX}}}

\copyrightyear{2022}
\acmYear{2022}
\setcopyright{acmcopyright}
\acmConference[CIKM '22]{Proceedings of the 31st ACM
International Conference on Information and Knowledge Management}{October
17--21, 2022}{Atlanta, GA, USA}
\acmBooktitle{Proceedings of the 31st ACM International Conference on Information
and Knowledge Management (CIKM '22), October 17--21, 2022, Atlanta, GA, USA}
\acmPrice{15.00}
\acmDOI{10.1145/3511808.3557713}
\acmISBN{978-1-4503-9236-5/22/10}

\usepackage{booktabs}
\usepackage{multirow}
\usepackage{xcolor}
\usepackage{dsfont}
\usepackage{adjustbox}
\usepackage{subfig}
\usepackage{enumitem}
\usepackage{xspace}

\definecolor{blue(pigment)}{rgb}{0.2, 0.2, 0.6}

\newcommand{\eg}{e.g., }
\newcommand{\ie}{i.e., }

\newcommand{\partitle}[1]{\vspace{2mm}\noindent\textbf{#1}}

\newcommand{\uga}{active\xspace}
\newcommand{\ugb}{semi-active\xspace}
\newcommand{\ugc}{inactive\xspace}

\newcommand{\ourmodel}{CCL\xspace}

\begin{document}

\title{Towards Confidence-aware Calibrated Recommendation}


\author{Mohammadmehdi Naghiaei}
\affiliation{%
  \institution{University of Southern California}
  \city{California}
  \country{USA}}
\email{naghaei@usc.edu}

\author{Hossein A.~Rahmani}
\affiliation{%
  \institution{University College London}
  \city{London}
  \country{UK}}
\email{h.rahmani@ucl.ac.uk}

\author{Mohammad Aliannejadi}
\affiliation{%
  \institution{University of Amsterdam}
  \city{Amsterdam}
  \country{The Netherlands}}
\email{m.aliannejadi@uva.nl}

\author{Nasim Sonboli}
\affiliation{%
  \institution{Tufts University}
  \city{Medford, MA}
  \country{USA}}
\email{nasim.sonboli@tufts.edu}

\renewcommand{\shortauthors}{M.~Naghiaei, H.~A.~Rahmani, M.~Aliannejadi, N.~Sonboli}

\begin{abstract}
Recommender systems utilize users' historical data to learn and predict their future interests, providing them with suggestions tailored to their tastes. Calibration ensures that the distribution of recommended item categories is consistent with the user's historical data. Mitigating miscalibration brings various benefits to a recommender system. For example, it becomes less likely that a system overlooks categories with less interaction on a user's profile by only recommending popular categories. Despite the notable success, calibration methods have several drawbacks, such as limiting the diversity of the recommended items and not considering the calibration confidence. This work, presents a set of properties that address various aspects of a desired calibrated recommender system. Considering these properties, we propose a confidence-aware optimization-based re-ranking algorithm to find the balance between calibration, relevance, and item diversity, while simultaneously accounting for calibration confidence based on user profile size. Our model outperforms state-of-the-art methods in terms of various accuracy and beyond-accuracy metrics for different user groups.
\end{abstract}



\maketitle

\section{Introduction}
\label{sec:intro}

Miscalibration over categories occurs when user profiles contain a different distribution of categories than the recommendation list. The calibration of user-profiles addresses this issue by maintaining the same distribution. Therefore, less popular genres are more likely to appear on the recommendation list. While the machine learning community has addressed calibration extensively~\cite{bella2010calibration}, it has recently gained attention in the recommender system community~\cite{steck2018calibrated,seymen2021constrained,Kun2020Calib}. \citet{steck2018calibrated} propose an iterative greedy algorithm to add the most relevant items to the recommendation list until they satisfy the calibration constraint. \citet{seymen2021constrained} argue that this approach is suboptimal because of the myopic nature of greedy algorithms. Instead, they propose an optimization algorithm for calibration while maintaining the overall effectiveness of the model.



The previous studies on calibrating recommendations aim for fair treatment of all. However, they mainly assume that the algorithm's capacity for calibration is the same for all users regardless of their profile sizes. For instance, user A with a profile size of 100 items is treated the same as user B with a profile size of 10.
We argue that the confidence in the categorical distribution of a user profile is proportional to one's profile size. Assume that users A and B have 9:1 proportions of comedy movies in their profiles compared to action movies in a domain with only two categories. Figure~\ref{fig:beta_dist} shows the beta distribution\footnote{With a Bayesian view of the two user profiles, we let $\theta_A$ and $\theta_B$ be the probability of watching a comedy show during test time, for users A and B, respectively. Therefore, our posterior for $\theta_A$ is $beta(91,11)$ and for $\theta_B$ is $beta(10,2)$.} of both users watching a comedy show in the test set ($\theta_A = beta(91,11)$ vs.~$\theta_B = beta(10,2)$). We observe user A is ~0.67 more likely to watch a comedy show, indicating the critical role of the profile size in defining calibration confidence. We also demonstrate the same result via empirical analysis in Figure~\ref{fig:testvstrain_diff_ml_100K} where we observe a higher miscalibration between the test and train sets as the user profile size decreases (e.g., active vs.~inactive users).

In this work, we resolve this issue by proposing a confidence-aware optimization-based re-ranking algorithm to find the balance between calibration and item diversity while maintaining distinct confidence levels in modeling calibration for different users. Inspired by our theoretical analysis, we propose a weighting scheme to model calibration confidence in an optimization framework capable of addressing user profiles' diverse and dynamic nature. Our extensive experiments demonstrate our method's effectiveness in terms of accuracy and beyond-accuracy metrics such as diversity. We further analyze the performance of our model for different user groups, showing our weighting scheme's effectiveness for users of different profile sizes. Our contributions are summarized as follows:



\begin{itemize}[leftmargin=*]
    \item Proposing a weighted mixed-integer linear programming model, called \ourmodel, to solve our multi-objective calibration problem.
    \item Evaluating the proposed method against the state-of-the-art recommendation calibration method applied on \textit{four} baseline recommendation algorithms and \textit{two} real-world datasets.
    \item Extensive empirical analysis of the results on various beyond-accuracy metrics such as diversity across \textit{three} user groups.\footnote{\url{https://github.com/rahmanidashti/CalibrationFair}}
\end{itemize}
\section{Calibrated Recommendation}
\label{sec:method}

\subsection{Problem Definition}
\label{sec:problem_definition}
Let $\mathcal{U}$ and $\mathcal{I}$ be the set of consumers and items with sizes $n$ and $m$, respectively. Further, assume that there exists a set of categories $\mathcal{C}$ describing items in $\mathcal{I}$.  Categories can be either defined by the system designer, such as popular and unpopular items~\cite{abdollahpouri2020connection}, or intrinsic features of items, such as a movie genre (e.g., Action, Comedy).  In a typical recommender system, a user, $u \in \mathcal{U}$, is provided a list of top-N items, $L_N(u)$ according to a defined relevance score, $s \in S^{n\times N}$. To measure the degree of calibration of the recommendation list, following~\citet{steck2018calibrated}, we determine two distributions for the categories $c \in \mathcal{C}$; $p(c|u)$, for the rated items in a user profile, and $q(c|u)$ for the items in her recommendation list $L_N(u)$. These distributions are formally defined as:


\begin{small}
\begin{equation}
\label{eq:p}
p(c|u) = \frac{\sum_{i \in \mathcal{I}_u} w_{ui} \times \mathds{1}(c \in i) }{|\mathcal{I}_u| \times \sum_{i \in \mathcal{I}_u} w_{ui}}
\end{equation}
\begin{equation}
\label{eq:q}
q(c|u) = \frac{\sum_{i \in L_N(u)} w_{r(i)} \times \mathds{1}(c \in i) }{|L_N(u)| \times \sum_{i \in L_N(u)} w_{r(i)}}
\end{equation}
\end{small}

\noindent
where $\mathds{1}(c \in i)$ is an indicator function returning 1 when item $i$ has category $c$ and 0 otherwise. $\mathcal{I}_u$ is the set of items in the profile of user $u$. $w_{r(i)}$ is the weight of item $i$ given its rank, $r(i)$,  in the recommendation list (\eg logarithmic reduction factor based on rank similar to nDCG's weighting scheme~\cite{wang2013theoretical}) whereas $w_{ui}$ represents the weight of item $i$ for user $u$ (\eg based on how recently the item has been rated). Finally, similar to~\cite{steck2018calibrated}, the degree of miscalibration is calculated by directly computing the statistical distance between two distributions of $p(g|u)$ and $q(g|u)$. In this work, we adopt Hellinger distance defined as \(H(p,q) = \frac{|| \sqrt{p} - \sqrt{q}||_{2}}{\sqrt{2}}\) as well as Jensen-Shannon (JS) divergence to calculate a symmetric normalized score using KL divergence. The smaller values of JS and H indicate better performance in calibration, where the perfect calibration occurs when they are equal to zero. The overall miscalibration metric, $\mathbf{MC}$, is calculated by averaging divergence values (\eg using JS or H functions) across all users.




\subsection{Desired Properties of Calibration}
\label{sec:property}

\begin{figure}
    \centering
    \subfloat[Beta PDF]
    {
        {
            \includegraphics[width=0.4\columnwidth]{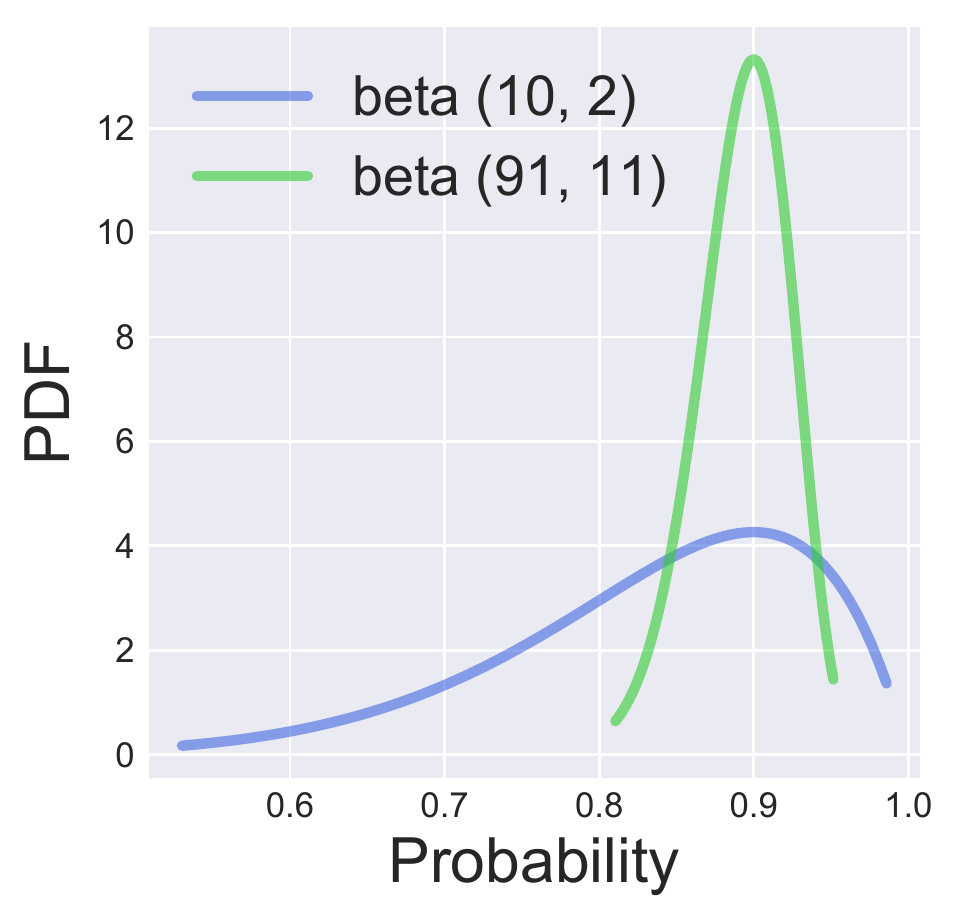}
            \label{fig:beta_dist}%
        }
    }
    \subfloat[Miscalibration in Test vs.~Train]
    {
        {
            \includegraphics[width=0.4\columnwidth]{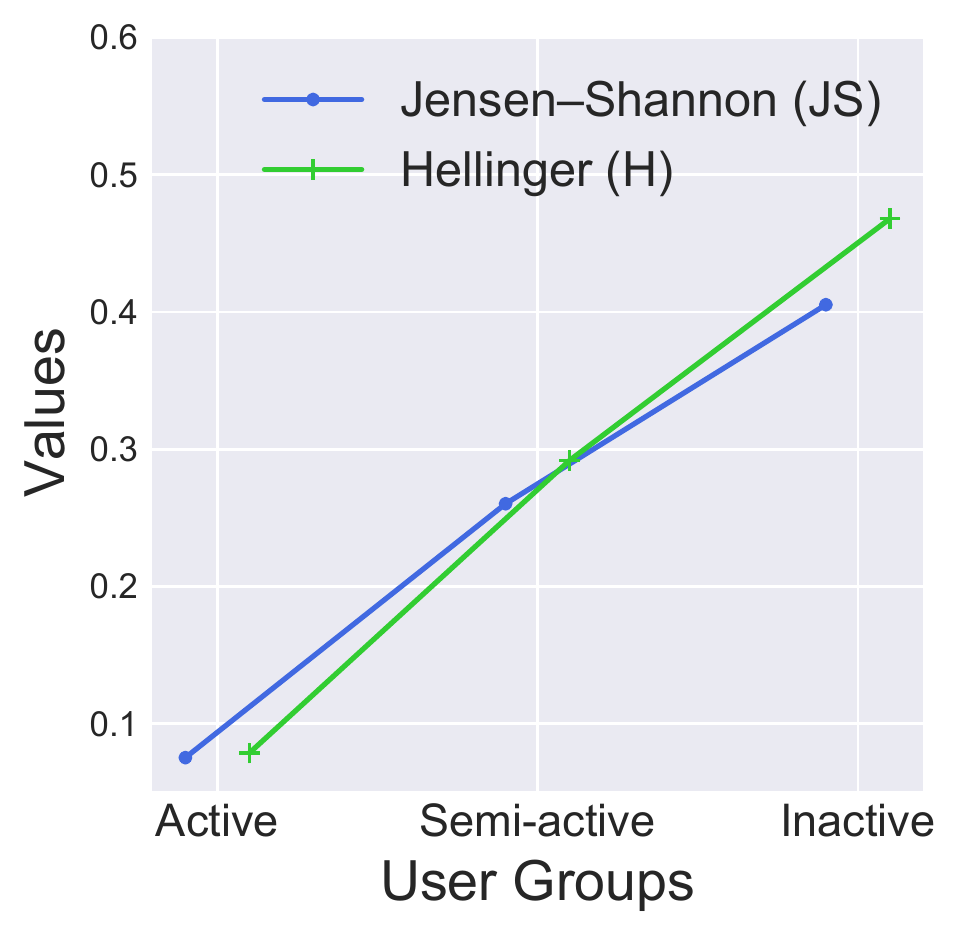}
            \label{fig:testvstrain_diff_ml_100K}
        }
    }
    \caption{Calibration for users with different profile size}
    \label{fig:motivation}%
\end{figure}

Here we present the desired properties of calibrated recommendations and explain how they affect the final recommendation before moving into our proposed model addressing them.

\partitle{P1: Relevance Property.}
Calibrated recommendations intend to reflect different interests of users, as indicated by their interaction history, with the same proportion in the recommended list. A greedy algorithm is often proposed to re-rank a baseline recommendation list. The first desired property is that calibrating categories should occur without sacrificing overall relevance to users. \citet{seymen2021constrained} illustrate that greedy calibrated algorithms such as \cite{steck2018calibrated, oh2011novel} fail to meet this essential requirement, leading to suboptimal recommendations.

\partitle{P2: Diversity Aspects.}
Diversity is defined in literature as minimal redundancy or similarity among the recommended items. Various research has been proposed to generate diverse recommendations~\cite{bhaskara2016linear,zhang2008avoiding,hurley2011novelty} . An alternative research direction focuses on achieving personalized diversity, i.e., recommendations should reflect the diverse interests of users with the correct proportion~\cite{dang2012diversity,vargas2014coverage,Eskandanian2017PersonalizedDiv,Sonb2020Opp}. A traditional calibrated recommendation follows the second approach and does not consider diversity as a separate objective that might lead to a so-called "filter bubble"~\cite{pariser2011filter}. Imagine a user rated 70\% Comedy and 30\% romance movies in a world with three genres. Such users would receive the target distribution of 70\%:30\%:0\% of the same genres in the calibrated recommended list, depriving them of the opportunity to interact with the third genre. The second desired property in calibrated recommendation is to prevent such phenomenon by preventing a reduction in overall catalog coverage and diversity in comparison to baselines. 


\partitle{P3: Calibration Confidence.}
Calibrated recommendations are solely based on users' historical interactions. As discussed earlier, user profile sizes influence the confidence level of the calibration algorithm. Moreover, this can be viewed from a temporal perspective, where users' tastes change over time. The more users spend time in a system interacting with items, the more we know about their future interests and vice versa. To investigate this issue, along with the theoretical analysis of the Beta distributions (see Section~\ref{sec:intro}), we perform a temporal split of the data into train/test scenarios. For this example, we consider the test data as an oracle recommendation algorithm that can make perfect predictions, \ie return the ground truth. Figure~\ref{fig:testvstrain_diff_ml_100K} shows the degree of miscalibration for the oracle algorithm as defined in Section~\ref{sec:problem_definition}. To analyze the miscalibration effect from the user's perspective, we categorize users into three groups based on their activity level, \ie number of interactions. The top 20$\%$ of users are categorized as \uga~group while the bottom 20$\%$ are assigned \ugc~group, and the remainder is categorized as \ugb~group. As we can see from the values of JS and H in Figure~\ref{fig:testvstrain_diff_ml_100K}, the accuracy-optimal oracle algorithm fails to calibrate the recommendation list according to genre distribution, especially for the \ugc~group followed by the \ugb~group. This result indicates modeling the genre preference of less active users is more challenging. Considerable training data is needed to capture users' tastes in categories with high confidence. This is also evident in Figure~\ref{fig:beta_dist}, where we see a ~0.67 higher confidence when modeling users with larger profile sizes. Hence, calibrating recommendations for all users to the same degree regardless of their profile size leads to undesired recommendations. This raises the need for calibrated recommendations that perform better for less active users. Hence, the third desired property for calibration algorithms is varying degrees of calibration for less active users to benefit from the baseline algorithm's predictive ability.

\subsection{Proposed Model}
In the following, we describe our proposed confidence-aware calibration algorithm that meets the desired properties mentioned in Section~\ref{sec:property}. The purpose of the algorithm is to re-rank a baseline recommendation ranking list, $L_N(u)$, in a way that the new top-K list, $L_K^c(u)$ where $K<N$, would reduce miscalibration without affecting the recommendation lists' quality. To this end, we define binary matrix $Z=[Z_{ui}]_{n \times N}$ to denote whether item $i \in \mathcal{I}$ is recommended to user $u \in \mathcal{U}$ in $L_K^c(u)$. Hence, we can represent $L_K^c(u)$ as $[Z_{u1}, Z_{u2}, ..., Z_{uN}]$ when $\sum_{i=1}^{N}{Z_{ui}} = K$. Let $W(I_u) \in [0,1]$ denote the confidence weight given to user $u$ based on the rated items in her profile, $I_u$. The confidence weight aims to represent the degree of calibration for user $u$ where $1$ indicate that the model would fully calibrate the user's historical preference in the recommended list while $0$ indicates no calibration. We can formalize the calibration optimization problem given the binary decision matrix, $Z$, as:

\begin{small}
\begin{equation}
\begin{aligned}
\max_{Z_{ui}} \quad & \sum_{u \in \mathcal{U}}\sum_{i=1}^{N}{S_{ui} \cdot Z_{ui}} - \lambda_1 \cdot \mathbf{MC}(Z^1,\mathcal{U})   \\
\textrm{s.t.} \quad & \sum_{i=1}^{N}{Z_{ui}} = K \quad,  \sum_{i=1}^{N}{Z^1_{ui}} = int(W(I_u) \cdot K) & \forall u \\ 
& Z^1_{ui} + Z^2_{ui} = Z_{ui} \quad,  Z^1_{ui}~, Z^2_{ui}~, Z_{ui} \in{\{0,1\}} & \forall u,i 
\end{aligned}
\label{eq:optimization}
\end{equation}
\end{small}

\noindent
where $Z^1_{ui}$ and $Z^2_{ui}$ are helper binary decision variables that indicate whether the item $Z_{ui}$ is selected due to improving miscalibration metric or a high predictive relevance score, respectively. The $int(.)$ function rounds the input to the nearest integer value. In our experiment, we define $W(I_u)$ based on the users' profile size, formally as $W(I_u) = \min [\frac{|I_u|}{\frac{1}{\mathcal{|U|}}  \sum_{u\in \mathcal{U}}|I_u|}, 1]$ motivated by our analysis in Figure~\ref{fig:testvstrain_diff_ml_100K} and leave more rigorous definitions for future work. Additionally, as described in Section~\ref{sec:problem_definition}, $\mathbf{MC}$ represents the miscalibration metric by calculating the average divergence values for all users on a subset of the top-K list by size $int(W(I_u) \cdot K)$ denoted by $Z^1$ decision matrix. Given that the JS and H divergence functions are non-linear, for optimization purposes we adopt the \textit{Total Variation} function~\cite{chambolle2004algorithm,levin2017markov} defined as the $\mathscr{l}_1$ norm between distributions of $p(c|u)$ and $q(c|u)$ to calculate divergence values. As a result of the linear objective function, the optimization problem becomes mixed-integer linear programming (MILP) that can be solved using heuristic algorithms (e.g., branch\&bound algorithms) in industrial optimization solvers to provide a satisfactory solution in practice\footnote{In our experiment, we used the Gurobi solver at \url{https://www.gurobi.com}}. 
Note that our proposed model will obviously satisfy properties \textbf{P1} and \textbf{P2} when $\lambda_1 = 0$ as it will return the same baseline recommendation and will calibrate the recommendation list completely for all users when $W(I_u)= 1 ~\forall u$ and $\lambda_1$ is large. Varying weighting parameter $W(I_u)$ aims to satisfy the property \textbf{P3}.

\section{Experiments and Results}
\label{sec:results}

\begin{figure}
    \centering
    \includegraphics[scale=0.40]{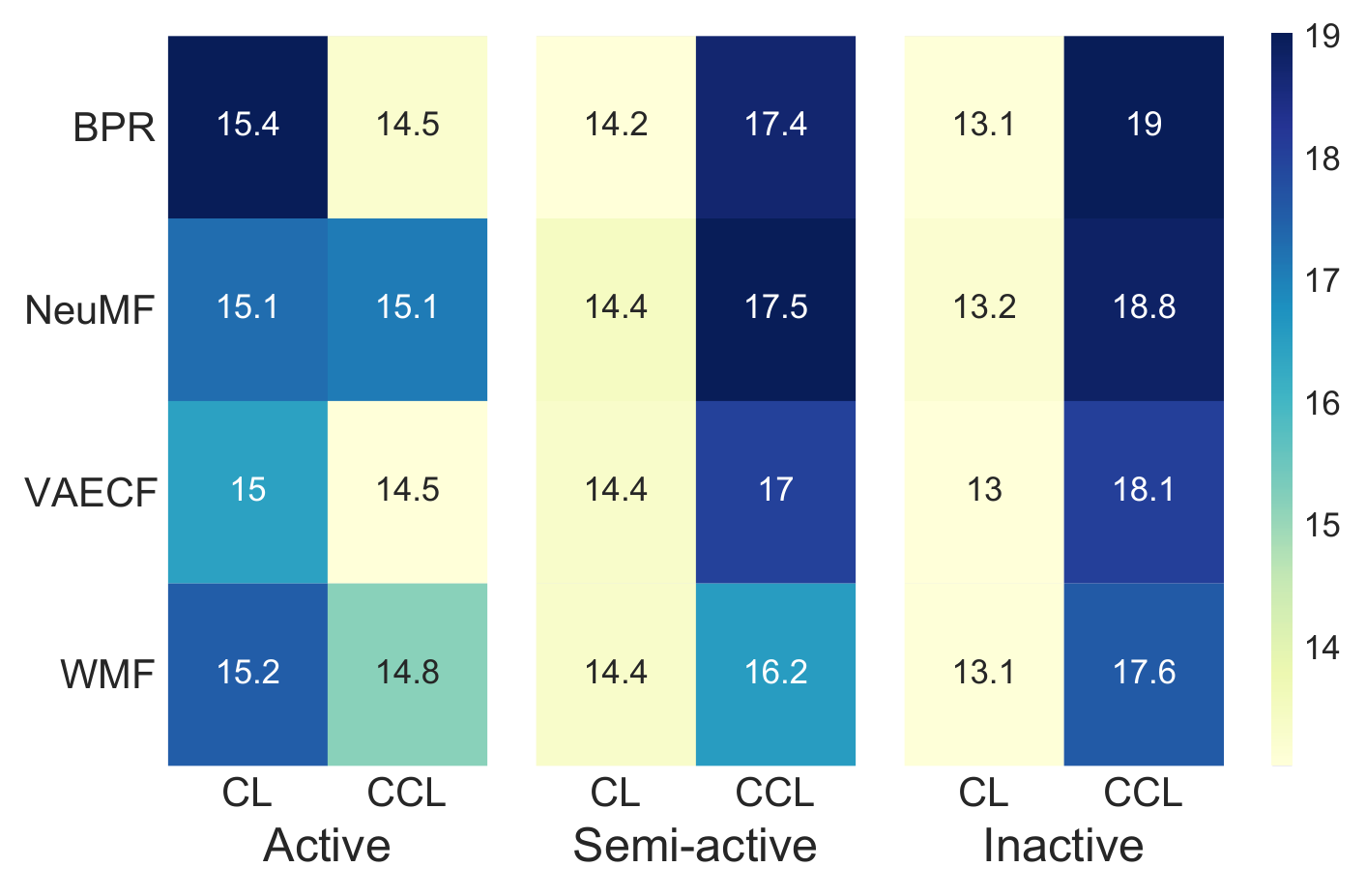}
    \caption{User groups' catalog coverage on different models.}
    \label{fig:heatmap_catcov_ml_100K}
\end{figure}

\begin{figure}
    \centering
    \subfloat[MovieLens1M]
    {
        {
            \includegraphics[width=0.49\columnwidth]{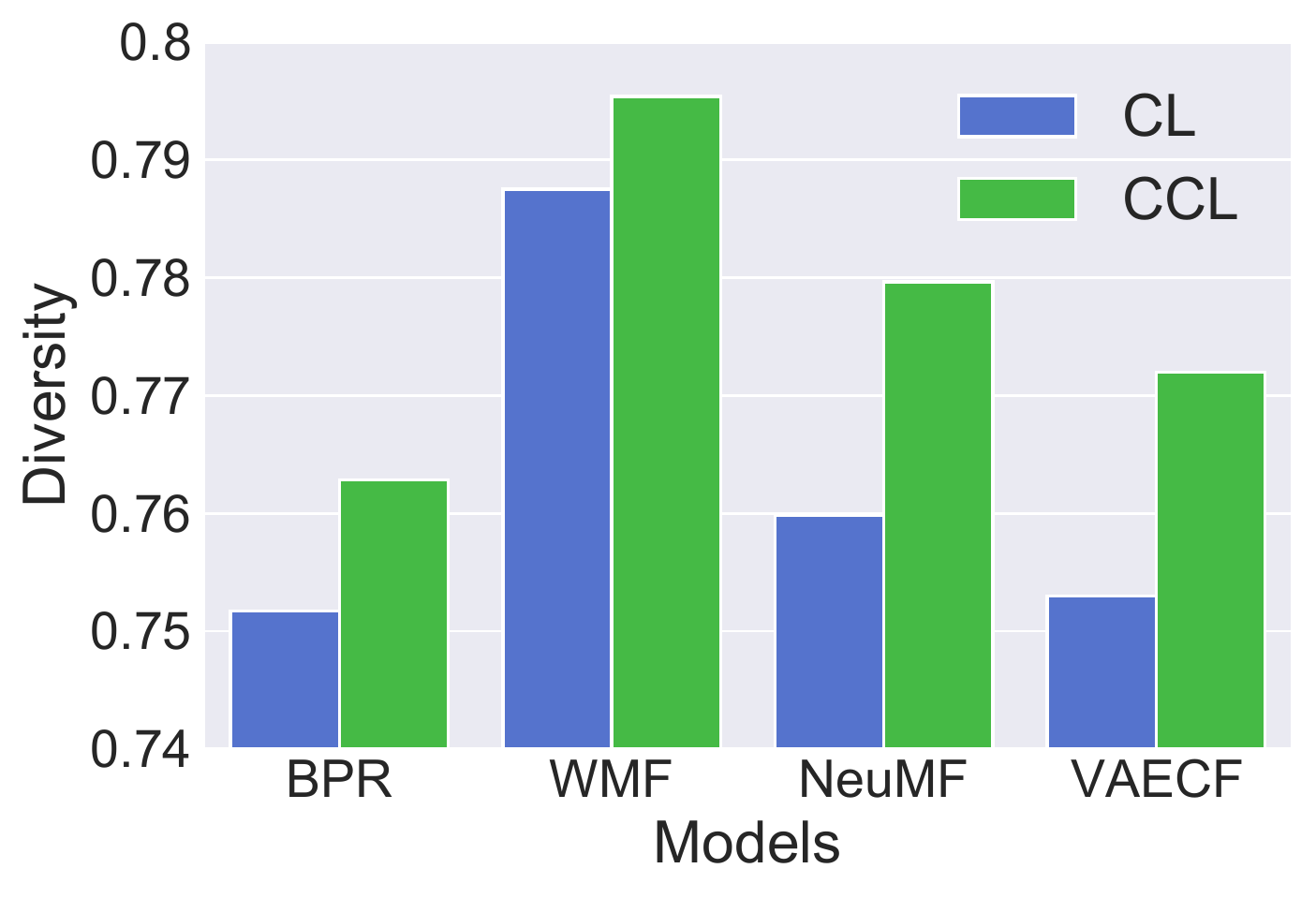}
        }
    }
    \subfloat[MovieLensSmall]
    {
        {
            \includegraphics[width=0.49\columnwidth]{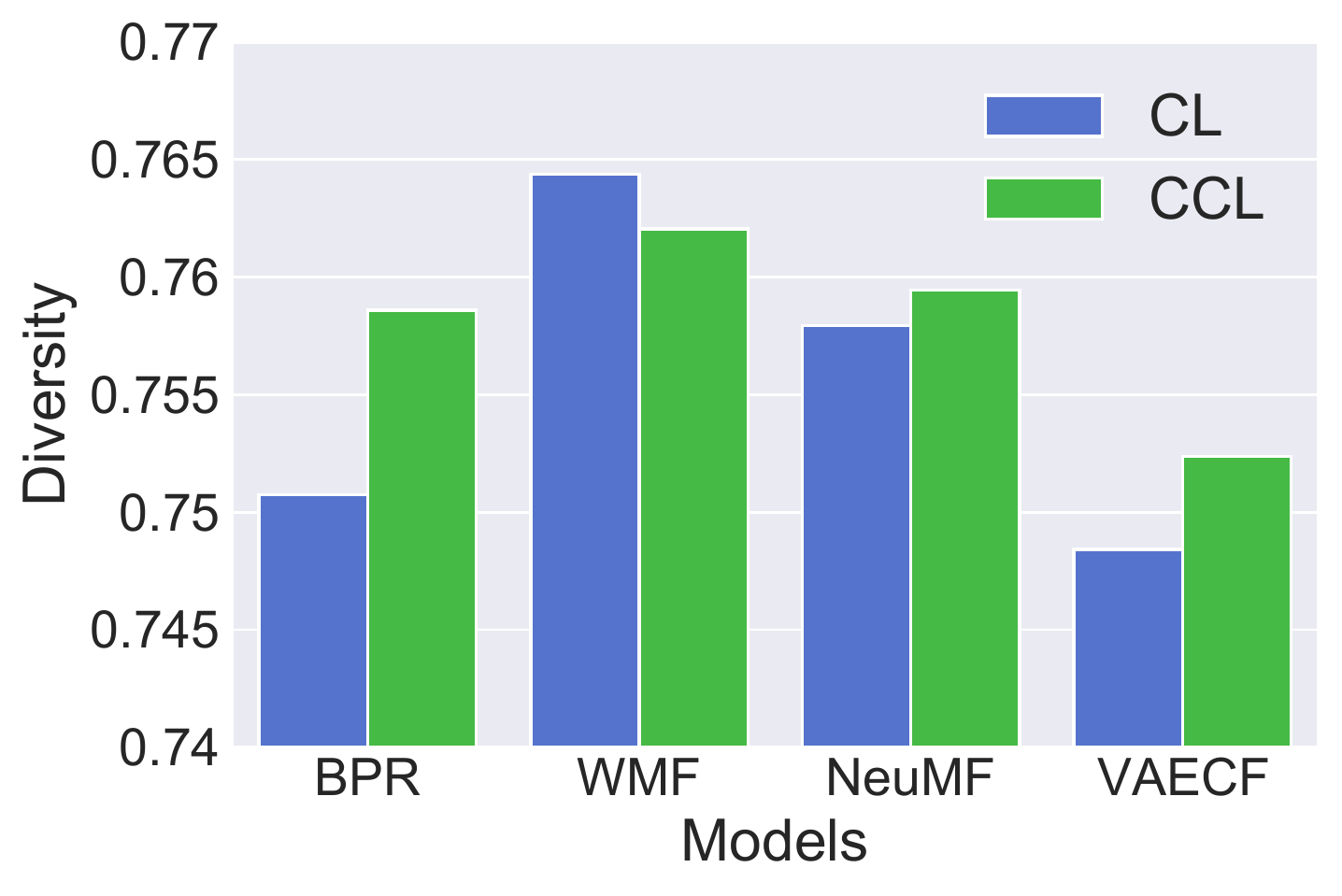}
        }
    }
    \caption{Diversity metric in CL and CCL model for baselines.}%
    \label{fig:diversity}%
\end{figure}

\begin{figure*}
    \centering
    \subfloat[nDCG]
    {
        {
            \includegraphics[scale=0.26]{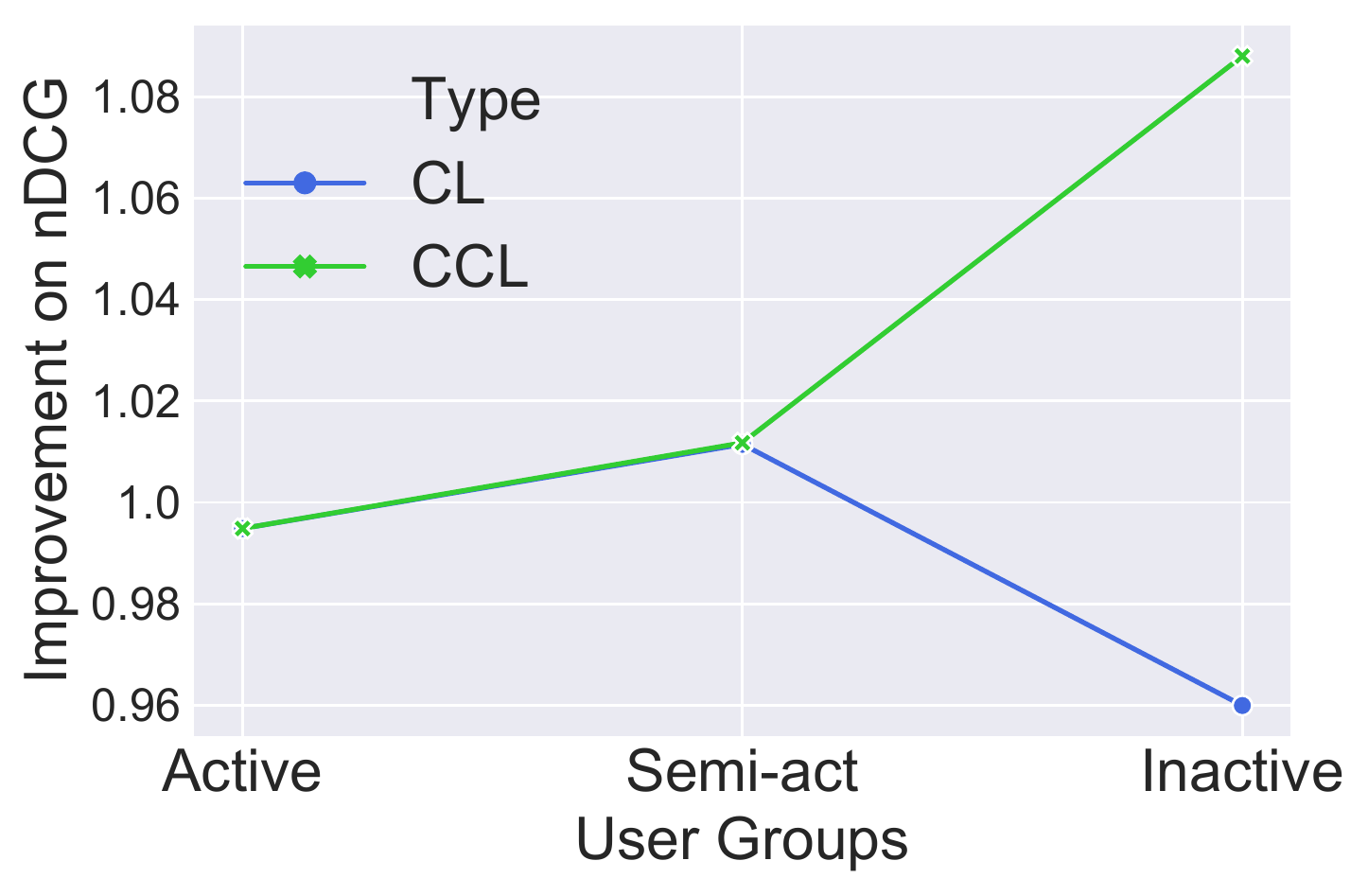}
        }
    }
    \subfloat[Miscalibration]
    {
        {
            \includegraphics[scale=0.26]{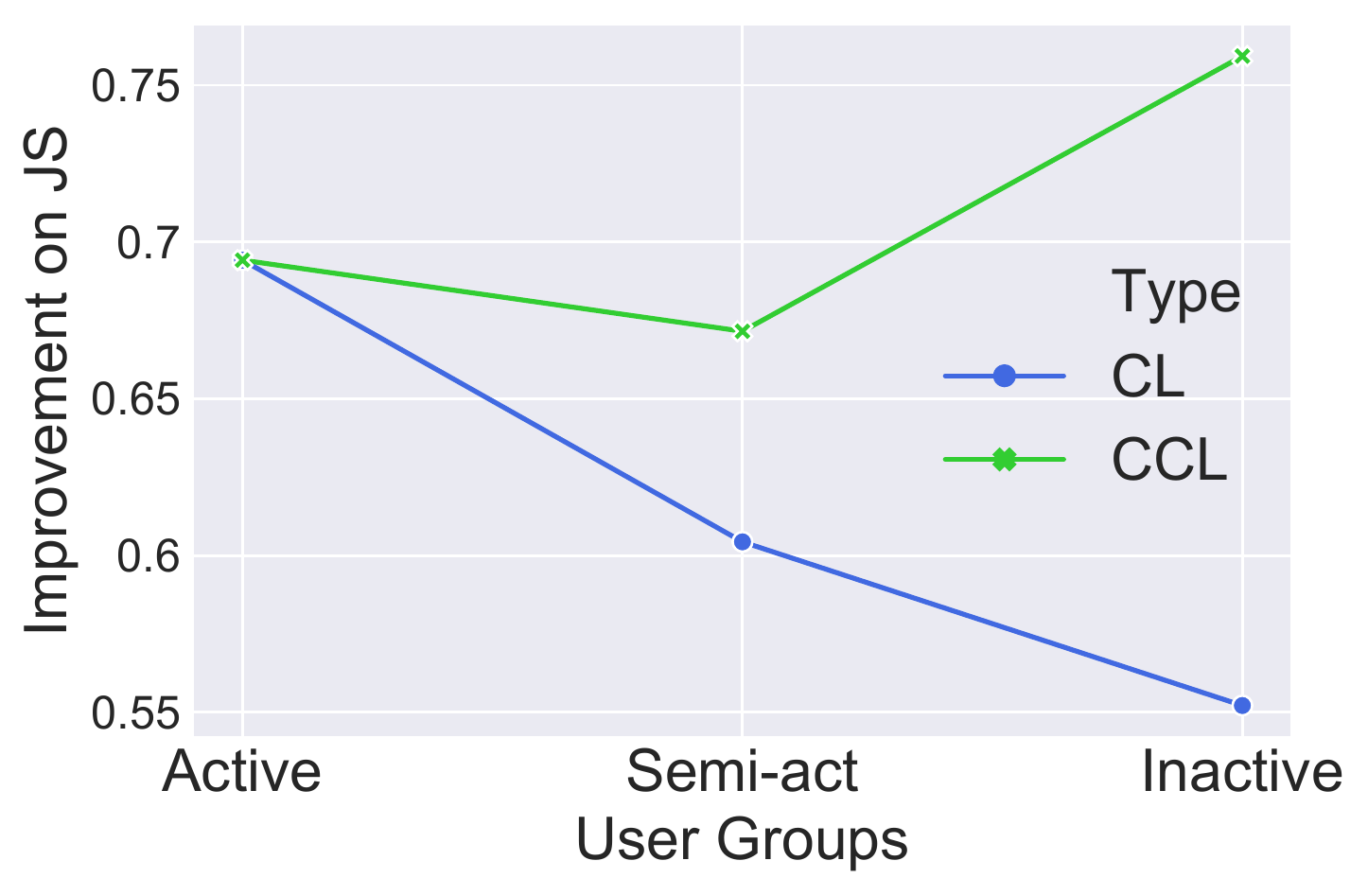}
        }
    }
    \subfloat[Diversity]
    {
        {
            \includegraphics[scale=0.26]{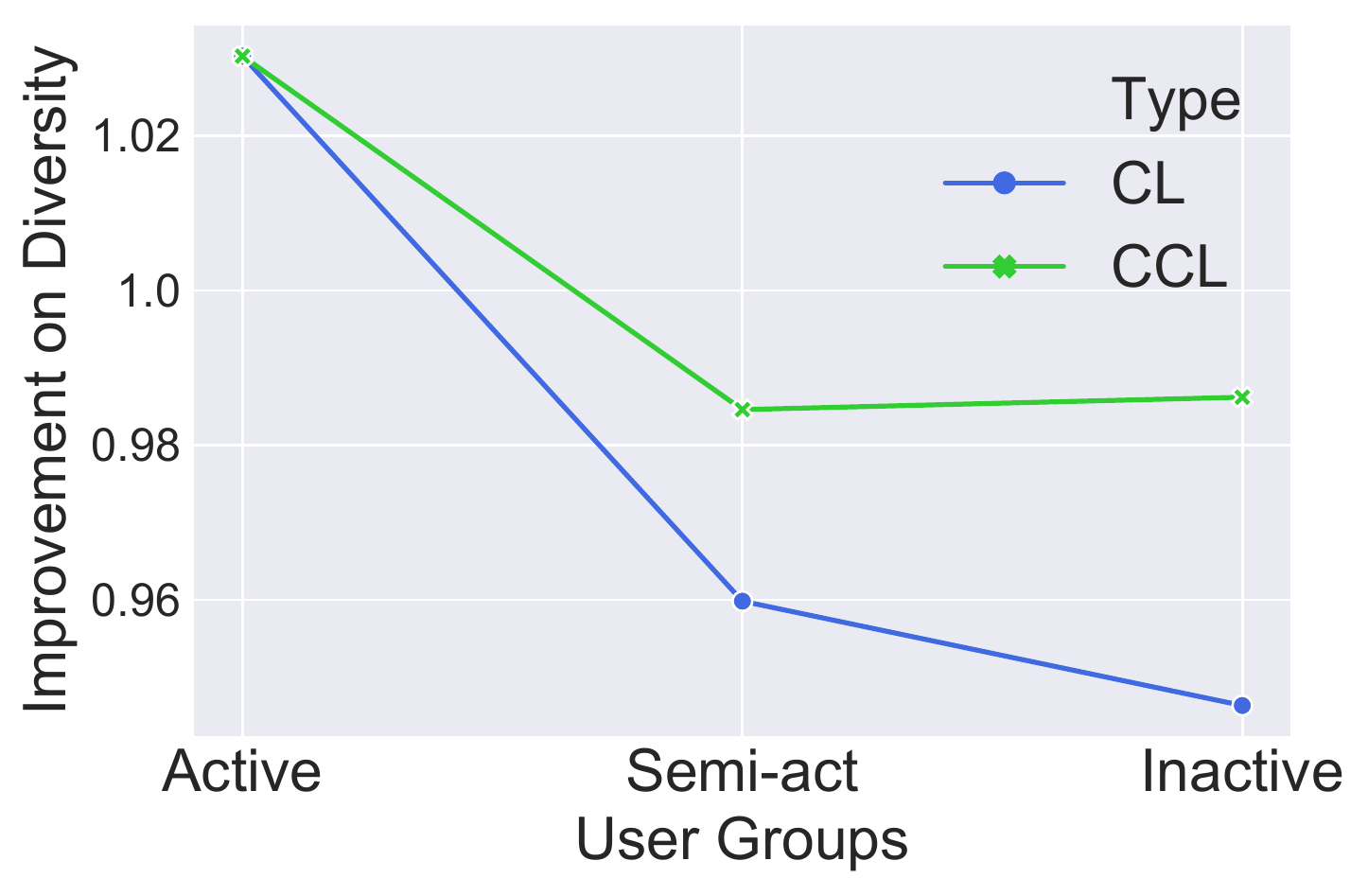}
        }
    }
    \subfloat[Catalog Coverage]
    {
        {
            \includegraphics[scale=0.26]{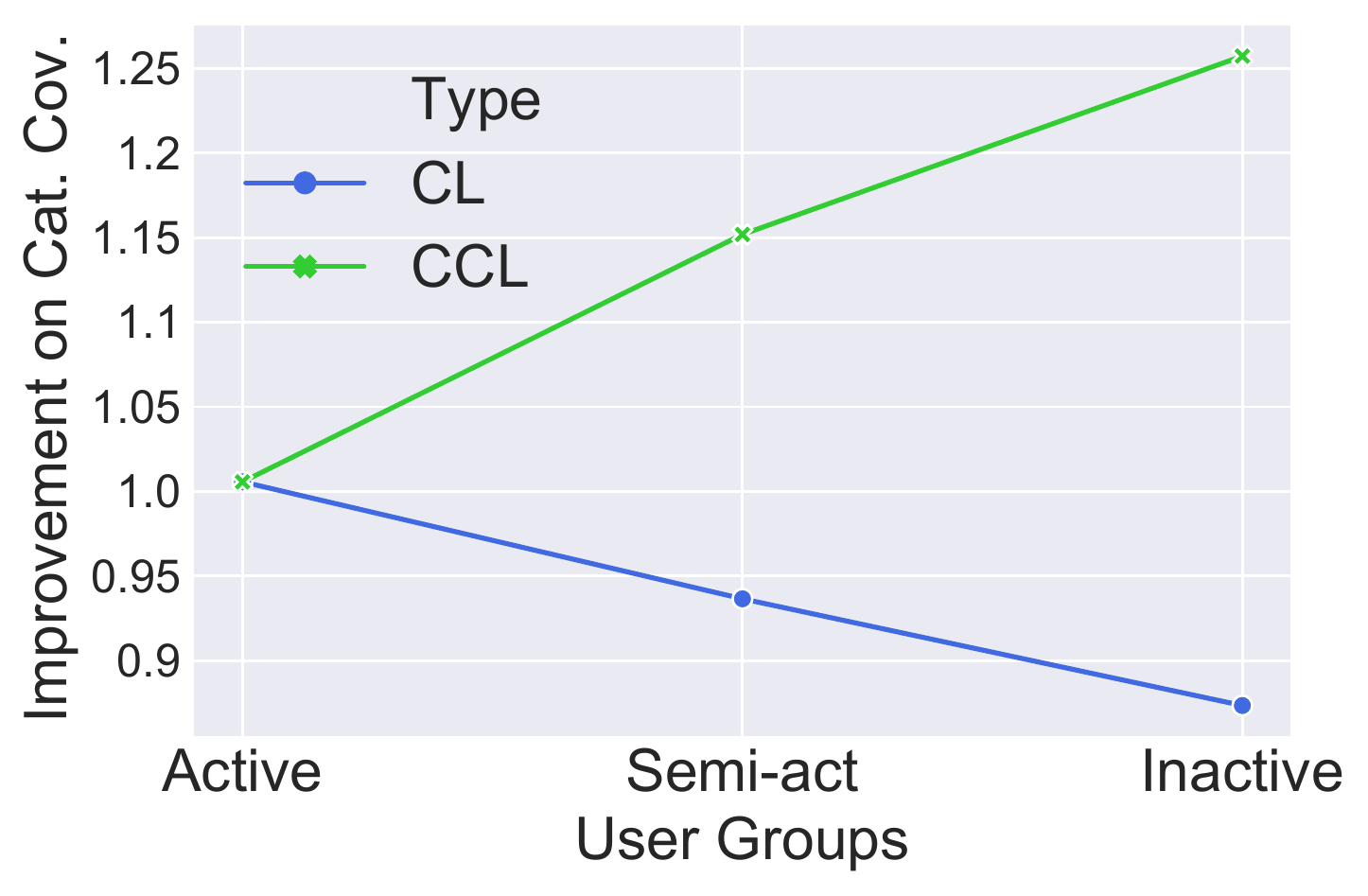}
        }
    }
    \caption{Improvement from the user's perspective over baselines on various performance aspects on MovieLens1M}%
    \label{fig:CCL_benefit}%
\end{figure*}

\begin{table*}
\centering
\caption{The recommendation performance of our re-ranking method and corresponding baselines on MovieLensSmall and MovieLens1M datasets. The evaluation metrics here are calculated based on the top-20 predictions in the test set. Our best results are highlighted in bold. $\dagger$ and $\ddagger$ signs show the t-test p-value significance at 0.05 compared to N and CL, respectively.}
\label{tab:comparison}
\resizebox{15cm}{!}{
\begin{adjustbox}{max width=\textwidth}
\begin{tabular}{llllllllllllllll}

\toprule
\multirow{2}{*}{Model} & \multirow{2}{*}{Type} && \multicolumn{6}{c}{MovieLensSmall} && \multicolumn{6}{c}{MovieLens1M} \\
\cline{4-9}\cline{11-16}
      &     && Precision    & Recall    & nDCG    & CatCov $\uparrow$ & JS $\downarrow$ & H $\downarrow$ && Precision & Recall & nDCG & CatCov $\uparrow$ & JS $\downarrow$ & H $\downarrow$ \\ \hline
\multirow{3}{*}{BPR}   & N   && 0.0605 & 0.0668 & 0.0650 & 14.133 & 0.290 & 0.318 && 0.0769 & 0.0645 & 0.0813 & 11.221 & 0.338 & 0.377 \\
   & CL  && 0.0607 & 0.0673 & 0.0668 & 14.193 & \textbf{0.130} & \textbf{0.144} && 0.0787 & 0.0664 & 0.0841 & 10.306 & \textbf{0.213} & \textbf{0.246} \\
   & CCL && \textbf{0.0630}$^{\dagger\ddagger}$ & \textbf{0.0712}$^{\dagger\ddagger}$ & \textbf{0.0693}$^{\dagger\ddagger}$ & \textbf{17.303}$^{\dagger\ddagger}$ & 0.221 & 0.244 && \textbf{0.0797}$^{\dagger}$ & \textbf{0.0689}$^{\dagger\ddagger}$ & \textbf{0.0853}$^{\dagger}$ & \textbf{12.637}$^{\dagger\ddagger}$ & 0.228 & 0.261 \\ \hline
\multirow{3}{*}{WMF}   & N   && 0.0622 & 0.0806 & 0.0667 & 14.929 & 0.275 & 0.302 && 0.0629 & 0.0583 & 0.0682 & 11.238 & 0.376 & 0.420 \\
   & CL  && 0.0626 & 0.0820 & 0.0651 & 14.331 & \textbf{0.125} & \textbf{0.140} && 0.0648 & 0.0597 & 0.0682 & 10.455 & \textbf{0.230} & \textbf{0.264} \\
   & CCL && \textbf{0.0652}$^{\dagger\ddagger}$ & \textbf{0.0847}$^{\dagger\ddagger}$ & \textbf{0.0693}$^{\dagger\ddagger}$ & \textbf{16.159}$^{\dagger\ddagger}$ & 0.210 & 0.232 && \textbf{0.0649}$^{\dagger}$ & \textbf{0.0601}$^{\dagger}$ & \textbf{0.0684} & \textbf{12.604}$^{\dagger\ddagger}$ & 0.243 & 0.278 \\ \hline
\multirow{3}{*}{NeuMF} & N   && 0.0659 & 0.0784 & 0.0721 & 14.911 & 0.282 & 0.311 && 0.1006 & 0.0979 & 0.1071 & 11.797 & 0.337 & 0.375 \\
 & CL  && 0.0675 & 0.0766 & \textbf{0.0731} & 14.399 & \textbf{0.128} & \textbf{0.142} && 0.0999 & 0.0987 & 0.1057 & 10.940 & \textbf{0.199} & \textbf{0.227} \\
 & CCL && \textbf{0.0682}$^{\dagger}$ & \textbf{0.0809}$^{\dagger\ddagger}$ & 0.0729 & \textbf{17.469}$^{\dagger\ddagger}$ & 0.209 & 0.231 && \textbf{0.1010} & \textbf{0.1013}$^{\dagger\ddagger}$ & \textbf{0.1074} & \textbf{13.485}$^{\dagger\ddagger}$ & 0.225 & 0.258 \\ \hline
\multirow{3}{*}{VAECF} & N   && 0.0587 & 0.0672 & 0.0655 & 13.883 & 0.294 & 0.322 && 0.0960 & 0.0935 & 0.1019 & 11.543 & 0.334 & 0.372 \\
 & CL  && 0.0627 & 0.0741 & 0.0703 & 14.323 & \textbf{0.136} & \textbf{0.150} && 0.0962 & 0.0946 & 0.1028 & 10.814 & \textbf{0.202} & \textbf{0.231} \\
 & CCL && \textbf{0.0634}$^{\dagger}$ & \textbf{0.0743}$^{\dagger}$ & \textbf{0.0713}$^{\dagger}$ & \textbf{16.903}$^{\dagger\ddagger}$ & 0.208 & 0.228 && \textbf{0.0966} & \textbf{0.0964}$^{\dagger\ddagger}$ & \textbf{0.1032} & \textbf{13.272}$^{\dagger\ddagger}$ & 0.226 & 0.259 \\ \hline
\bottomrule
\end{tabular}
\end{adjustbox}
}
\end{table*}

To this end, we compare our confidence-aware calibration algorithm (CCL) against baselines (N) and a variation of our model, (CL), where $W(I_u)=1, \forall u \in \mathcal{U}$ and hence the recommendation list is fully calibrated for all users. CL closely resembles to optimally tuned version of the prior work by~\citet{seymen2021constrained} for all users as a comparison with the state-of-the-art calibrated model. 

\partitle{Experimental Settings.} 
We use MovieLensSmall and MovieLens1M for our experiments. The MovieLensSmall contains $8974$ movies, $670$ users and $18$ item types (genres) and MovieLens1M includes $3260$ movies, $6040$ users and $18$ genres. Both datasets include explicit ratings between $1$ and $5$. We performed a temporal split on rating data into train and test sets in ratios of $80$\% and $20$\%, respectively. We estimate the rating matrix with BPR \cite{rendle2012bpr}, WMF \cite{hu2008collaborative}, NeuMF \cite{he2017neural}, and VAECF \cite{liang2018variational} implemented in Cornac\footnote{\url{https://cornac.readthedocs.io/}} which is a Python-based recommender system framework \cite{salah2020cornac,truong2021exploring} with their default hyperparameter in original paper. All baseline algorithms share the same train and test datasets. We treated the model parameter $\lambda_1$ as a hyperparameter and set them to the value that maximizes the  $nDCG / \mathbf{MC}$. We then compare the algorithms on Precision, Recall, nDCG, JS $\&$ H divergence (see Section~\ref{sec:problem_definition}), Diversity~\cite{shani2011evaluating}, and Catalog Coverage~\cite{vargas2014coverage} as evaluation metrics.

\partitle{Relevance Aspect.}
We report the recommendation performance in terms of Precision, Recall, and nDCG. As seen in Table~\ref{tab:comparison}, our proposed model achieves higher accuracy metrics than baseline models and CL, satisfying property \textbf{P1}. In the MovieLensSmall dataset, the average improvements in (Precision, Recall, nDCG) were (5\%, 6.1\%, 5\%) and (2.5\%, 3.70\%, 2.7\%) over N and CL, respectively. We observe a similar pattern in the Movielens1M dataset. In general, we observe that neural-based baseline models achieve higher accuracy metrics, with NeuMF having the best performance..


\partitle{Diversity Aspect.}
Figure~\ref{fig:diversity} represents the diversity of CCL and CL models on both datasets. When comparing the CCL and CL models, we realize that the CCL model increases the diversity of recommendations satisfying property \textbf{P2}. The magnitude of this increase differs between the two datasets, with a higher increase in MovieLens1M. Concerning baselines, we find the highest level of diversity in the WMF model, followed by NeuMF, indicating the inherent ability of such algorithms to incorporate relevant, diverse items in their top ranking list. 

Figure~\ref{fig:heatmap_catcov_ml_100K} depicts the catalog coverage of the CL and CCL models from the user's perspective. CCL model achieves significantly higher catalog coverage for inactive and semi-active groups. Furthermore, active user group catalog coverage remains relatively constant compared to CL. In avoiding potential filter bubbles, this result is desirable and achieves property \textbf{P2}. The reason is that the profile history of active users can often predict their taste toward categories (see Figure~\ref{fig:testvstrain_diff_ml_100K}), so these users benefit from recommendations from the categories they like to interact with. In contrast, the catalog coverage of the recommendation list for inactive users (newer users) is considerably expanded to expose them to various genres and alleviate filter bubbles. Users who are semi-active fall between the two extremes. 

\partitle{Calibration Confidence.}
Figure~\ref{fig:CCL_benefit} illustrates CL and CCL performance improvement compared to baselines from the user's perspective in terms of diversity, catalog coverage, nDCG, and miscalibration measured by JS divergence. For active users, in all aspects, CL and CCL models achieve similar and significant improvements compared to N. However, for inactive users, CCL leads to a higher degree of miscalibration, despite improving on all other measures. In other words, a higher degree of miscalibration in CCL (compared to CL) benefits inactive users by providing them with more relevant and diverse (measured by diversity and catalog coverage) recommendations. This emphasizes the challenge of recommendations for cold-start users. The performance of semi-active users falls in between these two categories. Hence, considering users' dynamic taste over time and not limiting baseline models' predictive abilities (i.e., following property \textbf{P3}) can benefit less active users, which our proposed CCL model captures.


\section{Conclusions and Future Work}
\label{sec:conclusion}
In this paper, we highlighted the desired properties in various aspects of calibration and motivated their importance. We then proposed a multi-objective mixed integer linear programming method to recommend a calibrated top-K recommendation. Extensive experiments on two datasets and four state-of-the-art recommendation algorithms confirm that our method can calibrate the recommendation list and simultaneously increase diversity and accuracy while accounting for calibration confidence. In the future, we plan to study calibration on various beyond-accuracy aspects, such as calibrating over degrees of interest towards popular items.

\begin{acks}
Author Sonboli's effort was supported by the CRA CIFellows program funded by the National Science Foundation under Grant No.~2127309.
\end{acks}

\bibliographystyle{ACM-Reference-Format}
\bibliography{reference}


\begin{thebibliography}{24}


\ifx \showCODEN    \undefined \def \showCODEN     #1{\unskip}     \fi
\ifx \showDOI      \undefined \def \showDOI       #1{#1}\fi
\ifx \showISBNx    \undefined \def \showISBNx     #1{\unskip}     \fi
\ifx \showISBNxiii \undefined \def \showISBNxiii  #1{\unskip}     \fi
\ifx \showISSN     \undefined \def \showISSN      #1{\unskip}     \fi
\ifx \showLCCN     \undefined \def \showLCCN      #1{\unskip}     \fi
\ifx \shownote     \undefined \def \shownote      #1{#1}          \fi
\ifx \showarticletitle \undefined \def \showarticletitle #1{#1}   \fi
\ifx \showURL      \undefined \def \showURL       {\relax}        \fi
\providecommand\bibfield[2]{#2}
\providecommand\bibinfo[2]{#2}
\providecommand\natexlab[1]{#1}
\providecommand\showeprint[2][]{arXiv:#2}

\bibitem[Abdollahpouri et~al\mbox{.}(2020)]%
        {abdollahpouri2020connection}
\bibfield{author}{\bibinfo{person}{Himan Abdollahpouri},
  \bibinfo{person}{Masoud Mansoury}, \bibinfo{person}{Robin Burke}, {and}
  \bibinfo{person}{Bamshad Mobasher}.} \bibinfo{year}{2020}\natexlab{}.
\newblock \showarticletitle{The connection between popularity bias,
  calibration, and fairness in recommendation}. In
  \bibinfo{booktitle}{\emph{Fourteenth ACM conference on recommender systems}}.
  \bibinfo{pages}{726--731}.
\newblock


\bibitem[Bella et~al\mbox{.}(2010)]%
        {bella2010calibration}
\bibfield{author}{\bibinfo{person}{Antonio Bella}, \bibinfo{person}{C{\`e}sar
  Ferri}, \bibinfo{person}{Jos{\'e} Hern{\'a}ndez-Orallo}, {and}
  \bibinfo{person}{Mar{\'\i}a~Jos{\'e} Ram{\'\i}rez-Quintana}.}
  \bibinfo{year}{2010}\natexlab{}.
\newblock \showarticletitle{Calibration of machine learning models}.
\newblock In \bibinfo{booktitle}{\emph{Handbook of Research on Machine Learning
  Applications and Trends: Algorithms, Methods, and Techniques}}.
  \bibinfo{publisher}{IGI Global}, \bibinfo{pages}{128--146}.
\newblock


\bibitem[Bhaskara et~al\mbox{.}(2016)]%
        {bhaskara2016linear}
\bibfield{author}{\bibinfo{person}{Aditya Bhaskara}, \bibinfo{person}{Mehrdad
  Ghadiri}, \bibinfo{person}{Vahab Mirrokni}, {and} \bibinfo{person}{Ola
  Svensson}.} \bibinfo{year}{2016}\natexlab{}.
\newblock \showarticletitle{Linear relaxations for finding diverse elements in
  metric spaces}.
\newblock \bibinfo{journal}{\emph{Advances in neural information processing
  systems}}  \bibinfo{volume}{29} (\bibinfo{year}{2016}).
\newblock


\bibitem[Chambolle(2004)]%
        {chambolle2004algorithm}
\bibfield{author}{\bibinfo{person}{Antonin Chambolle}.}
  \bibinfo{year}{2004}\natexlab{}.
\newblock \showarticletitle{An algorithm for total variation minimization and
  applications}.
\newblock \bibinfo{journal}{\emph{Journal of Mathematical imaging and vision}}
  \bibinfo{volume}{20}, \bibinfo{number}{1} (\bibinfo{year}{2004}),
  \bibinfo{pages}{89--97}.
\newblock


\bibitem[Dang and Croft(2012)]%
        {dang2012diversity}
\bibfield{author}{\bibinfo{person}{Van Dang} {and} \bibinfo{person}{W~Bruce
  Croft}.} \bibinfo{year}{2012}\natexlab{}.
\newblock \showarticletitle{Diversity by proportionality: an election-based
  approach to search result diversification}. In
  \bibinfo{booktitle}{\emph{Proceedings of the 35th international ACM SIGIR
  conference on Research and development in information retrieval}}.
  \bibinfo{pages}{65--74}.
\newblock


\bibitem[Eskandanian et~al\mbox{.}(2017)]%
        {Eskandanian2017PersonalizedDiv}
\bibfield{author}{\bibinfo{person}{Farzad Eskandanian},
  \bibinfo{person}{Bamshad Mobasher}, {and} \bibinfo{person}{Robin Burke}.}
  \bibinfo{year}{2017}\natexlab{}.
\newblock \showarticletitle{A Clustering Approach for Personalizing Diversity
  in Collaborative Recommender Systems}. In
  \bibinfo{booktitle}{\emph{Proceedings of the 25th Conference on User
  Modeling, Adaptation and Personalization}} (Bratislava, Slovakia)
  \emph{(\bibinfo{series}{UMAP '17})}. \bibinfo{publisher}{Association for
  Computing Machinery}, \bibinfo{address}{New York, NY, USA},
  \bibinfo{pages}{280–284}.
\newblock
\showISBNx{9781450346351}
\urldef\tempurl%
\url{https://doi.org/10.1145/3079628.3079699}
\showDOI{\tempurl}


\bibitem[He et~al\mbox{.}(2017)]%
        {he2017neural}
\bibfield{author}{\bibinfo{person}{Xiangnan He}, \bibinfo{person}{Lizi Liao},
  \bibinfo{person}{Hanwang Zhang}, \bibinfo{person}{Liqiang Nie},
  \bibinfo{person}{Xia Hu}, {and} \bibinfo{person}{Tat-Seng Chua}.}
  \bibinfo{year}{2017}\natexlab{}.
\newblock \showarticletitle{Neural collaborative filtering}. In
  \bibinfo{booktitle}{\emph{Proceedings of the 26th international conference on
  world wide web}}. \bibinfo{pages}{173--182}.
\newblock


\bibitem[Hu et~al\mbox{.}(2008)]%
        {hu2008collaborative}
\bibfield{author}{\bibinfo{person}{Yifan Hu}, \bibinfo{person}{Yehuda Koren},
  {and} \bibinfo{person}{Chris Volinsky}.} \bibinfo{year}{2008}\natexlab{}.
\newblock \showarticletitle{Collaborative filtering for implicit feedback
  datasets}. In \bibinfo{booktitle}{\emph{2008 Eighth IEEE international
  conference on data mining}}. Ieee, \bibinfo{pages}{263--272}.
\newblock


\bibitem[Hurley and Zhang(2011)]%
        {hurley2011novelty}
\bibfield{author}{\bibinfo{person}{Neil Hurley} {and} \bibinfo{person}{Mi
  Zhang}.} \bibinfo{year}{2011}\natexlab{}.
\newblock \showarticletitle{Novelty and diversity in top-n
  recommendation--analysis and evaluation}.
\newblock \bibinfo{journal}{\emph{ACM Transactions on Internet Technology
  (TOIT)}} \bibinfo{volume}{10}, \bibinfo{number}{4} (\bibinfo{year}{2011}),
  \bibinfo{pages}{1--30}.
\newblock


\bibitem[Levin and Peres(2017)]%
        {levin2017markov}
\bibfield{author}{\bibinfo{person}{David~A Levin} {and} \bibinfo{person}{Yuval
  Peres}.} \bibinfo{year}{2017}\natexlab{}.
\newblock \bibinfo{booktitle}{\emph{Markov chains and mixing times}}.
  Vol.~\bibinfo{volume}{107}.
\newblock \bibinfo{publisher}{American Mathematical Soc.}
\newblock


\bibitem[Liang et~al\mbox{.}(2018)]%
        {liang2018variational}
\bibfield{author}{\bibinfo{person}{Dawen Liang}, \bibinfo{person}{Rahul~G
  Krishnan}, \bibinfo{person}{Matthew~D Hoffman}, {and} \bibinfo{person}{Tony
  Jebara}.} \bibinfo{year}{2018}\natexlab{}.
\newblock \showarticletitle{Variational autoencoders for collaborative
  filtering}. In \bibinfo{booktitle}{\emph{Proceedings of the 2018 world wide
  web conference}}. \bibinfo{pages}{689--698}.
\newblock


\bibitem[Lin et~al\mbox{.}(2020)]%
        {Kun2020Calib}
\bibfield{author}{\bibinfo{person}{Kun Lin}, \bibinfo{person}{Nasim Sonboli},
  \bibinfo{person}{Bamshad Mobasher}, {and} \bibinfo{person}{Robin Burke}.}
  \bibinfo{year}{2020}\natexlab{}.
\newblock \showarticletitle{Calibration in Collaborative Filtering Recommender
  Systems: A User-Centered Analysis}. In \bibinfo{booktitle}{\emph{Proceedings
  of the 31st ACM Conference on Hypertext and Social Media}} (Virtual Event,
  USA) \emph{(\bibinfo{series}{HT '20})}. \bibinfo{publisher}{Association for
  Computing Machinery}, \bibinfo{address}{New York, NY, USA},
  \bibinfo{pages}{197–206}.
\newblock
\showISBNx{9781450370981}
\urldef\tempurl%
\url{https://doi.org/10.1145/3372923.3404793}
\showDOI{\tempurl}


\bibitem[Oh et~al\mbox{.}(2011)]%
        {oh2011novel}
\bibfield{author}{\bibinfo{person}{Jinoh Oh}, \bibinfo{person}{Sun Park},
  \bibinfo{person}{Hwanjo Yu}, \bibinfo{person}{Min Song}, {and}
  \bibinfo{person}{Seung-Taek Park}.} \bibinfo{year}{2011}\natexlab{}.
\newblock \showarticletitle{Novel recommendation based on personal popularity
  tendency}. In \bibinfo{booktitle}{\emph{2011 IEEE 11th International
  Conference on Data Mining}}. IEEE, \bibinfo{pages}{507--516}.
\newblock


\bibitem[Pariser(2011)]%
        {pariser2011filter}
\bibfield{author}{\bibinfo{person}{Eli Pariser}.}
  \bibinfo{year}{2011}\natexlab{}.
\newblock \bibinfo{booktitle}{\emph{The filter bubble: What the Internet is
  hiding from you}}.
\newblock \bibinfo{publisher}{Penguin UK}.
\newblock


\bibitem[Rendle et~al\mbox{.}(2012)]%
        {rendle2012bpr}
\bibfield{author}{\bibinfo{person}{Steffen Rendle}, \bibinfo{person}{Christoph
  Freudenthaler}, \bibinfo{person}{Zeno Gantner}, {and} \bibinfo{person}{Lars
  Schmidt-Thieme}.} \bibinfo{year}{2012}\natexlab{}.
\newblock \showarticletitle{BPR: Bayesian personalized ranking from implicit
  feedback}.
\newblock \bibinfo{journal}{\emph{arXiv preprint arXiv:1205.2618}}
  (\bibinfo{year}{2012}).
\newblock


\bibitem[Salah et~al\mbox{.}(2020)]%
        {salah2020cornac}
\bibfield{author}{\bibinfo{person}{Aghiles Salah}, \bibinfo{person}{Quoc-Tuan
  Truong}, {and} \bibinfo{person}{Hady~W Lauw}.}
  \bibinfo{year}{2020}\natexlab{}.
\newblock \showarticletitle{Cornac: A Comparative Framework for Multimodal
  Recommender Systems}.
\newblock \bibinfo{journal}{\emph{Journal of Machine Learning Research}}
  \bibinfo{volume}{21}, \bibinfo{number}{95} (\bibinfo{year}{2020}),
  \bibinfo{pages}{1--5}.
\newblock


\bibitem[Seymen et~al\mbox{.}(2021)]%
        {seymen2021constrained}
\bibfield{author}{\bibinfo{person}{Sinan Seymen}, \bibinfo{person}{Himan
  Abdollahpouri}, {and} \bibinfo{person}{Edward~C Malthouse}.}
  \bibinfo{year}{2021}\natexlab{}.
\newblock \showarticletitle{A Constrained Optimization Approach for Calibrated
  Recommendations}. In \bibinfo{booktitle}{\emph{Fifteenth ACM Conference on
  Recommender Systems}}. \bibinfo{pages}{607--612}.
\newblock


\bibitem[Shani and Gunawardana(2011)]%
        {shani2011evaluating}
\bibfield{author}{\bibinfo{person}{Guy Shani} {and} \bibinfo{person}{Asela
  Gunawardana}.} \bibinfo{year}{2011}\natexlab{}.
\newblock \showarticletitle{Evaluating recommendation systems}.
\newblock In \bibinfo{booktitle}{\emph{Recommender systems handbook}}.
  \bibinfo{publisher}{Springer}, \bibinfo{pages}{257--297}.
\newblock


\bibitem[Sonboli et~al\mbox{.}(2020)]%
        {Sonb2020Opp}
\bibfield{author}{\bibinfo{person}{Nasim Sonboli}, \bibinfo{person}{Farzad
  Eskandanian}, \bibinfo{person}{Robin Burke}, \bibinfo{person}{Weiwen Liu},
  {and} \bibinfo{person}{Bamshad Mobasher}.} \bibinfo{year}{2020}\natexlab{}.
\newblock \showarticletitle{Opportunistic Multi-Aspect Fairness through
  Personalized Re-Ranking}. In \bibinfo{booktitle}{\emph{Proceedings of the
  28th ACM Conference on User Modeling, Adaptation and Personalization}}
  (Genoa, Italy) \emph{(\bibinfo{series}{UMAP '20})}.
  \bibinfo{publisher}{Association for Computing Machinery},
  \bibinfo{address}{New York, NY, USA}, \bibinfo{pages}{239–247}.
\newblock
\showISBNx{9781450368612}
\urldef\tempurl%
\url{https://doi.org/10.1145/3340631.3394846}
\showDOI{\tempurl}


\bibitem[Steck(2018)]%
        {steck2018calibrated}
\bibfield{author}{\bibinfo{person}{Harald Steck}.}
  \bibinfo{year}{2018}\natexlab{}.
\newblock \showarticletitle{Calibrated recommendations}. In
  \bibinfo{booktitle}{\emph{Proceedings of the 12th ACM conference on
  recommender systems}}. \bibinfo{pages}{154--162}.
\newblock


\bibitem[Truong et~al\mbox{.}(2021)]%
        {truong2021exploring}
\bibfield{author}{\bibinfo{person}{Quoc-Tuan Truong}, \bibinfo{person}{Aghiles
  Salah}, \bibinfo{person}{Thanh-Binh Tran}, \bibinfo{person}{Jingyao Guo},
  {and} \bibinfo{person}{Hady~W Lauw}.} \bibinfo{year}{2021}\natexlab{}.
\newblock \showarticletitle{Exploring Cross-Modality Utilization in Recommender
  Systems}.
\newblock \bibinfo{journal}{\emph{IEEE Internet Computing}}
  (\bibinfo{year}{2021}).
\newblock


\bibitem[Vargas et~al\mbox{.}(2014)]%
        {vargas2014coverage}
\bibfield{author}{\bibinfo{person}{Sa{\'u}l Vargas}, \bibinfo{person}{Linas
  Baltrunas}, \bibinfo{person}{Alexandros Karatzoglou}, {and}
  \bibinfo{person}{Pablo Castells}.} \bibinfo{year}{2014}\natexlab{}.
\newblock \showarticletitle{Coverage, redundancy and size-awareness in genre
  diversity for recommender systems}. In \bibinfo{booktitle}{\emph{Proceedings
  of the 8th ACM Conference on Recommender systems}}.
  \bibinfo{pages}{209--216}.
\newblock


\bibitem[Wang et~al\mbox{.}(2013)]%
        {wang2013theoretical}
\bibfield{author}{\bibinfo{person}{Yining Wang}, \bibinfo{person}{Liwei Wang},
  \bibinfo{person}{Yuanzhi Li}, \bibinfo{person}{Di He}, {and}
  \bibinfo{person}{Tie-Yan Liu}.} \bibinfo{year}{2013}\natexlab{}.
\newblock \showarticletitle{A theoretical analysis of NDCG type ranking
  measures}. In \bibinfo{booktitle}{\emph{Conference on learning theory}}.
  PMLR, \bibinfo{pages}{25--54}.
\newblock


\bibitem[Zhang and Hurley(2008)]%
        {zhang2008avoiding}
\bibfield{author}{\bibinfo{person}{Mi Zhang} {and} \bibinfo{person}{Neil
  Hurley}.} \bibinfo{year}{2008}\natexlab{}.
\newblock \showarticletitle{Avoiding monotony: improving the diversity of
  recommendation lists}. In \bibinfo{booktitle}{\emph{Proceedings of the 2008
  ACM conference on Recommender systems}}. \bibinfo{pages}{123--130}.
\newblock


\end{thebibliography}

\end{document}